# STELLA: A 16nm Spatio-Temporal Elastic Low-Latency CGRA for Multi-Stage Pipelined Applications

Jun Yin, Chao Fang*, Ryan Antonio, Xiaoling Yi, Yunhao Deng, Fanchen Kong, Marian Verhelst

ESAT-MICAS, KU Leuven, Belgium

*{jun.yin, chao.fang, ryan.antonio, xiaoling.yi, yunhao.deng, fanchen.kong, marian.verhelst}@kuleuven.be*

***Abstract* - Emerging non-matrix ML kernels, such as LayerNorm, GeLu, FFT or circular convolutions, demand low-latency, energy-efficient spatial accelerators beyond MatMul-centric arrays. STELLA presents a spatio-temporal elastic 16 nm coarse-grained reconfigurable array (CGRA) with a rapid configuration path, per-PE hardware loop control, and a low-latency, deeply pipelined elastic fabric with spatio-temporal data reuse. STELLA reaches up to 110 GOPS/mm$^2$ at 850 MHz, and improves effective kernel throughput by 4.84-7.14x over baseline CGRAs.**

**Keywords - CGRA, Elasticity, Spatio-Temporal, Irregular ML kernels.**

## I. Introduction

From traditional deep neural networks (DNNs) to Large Language Models (LLMs), AI algorithms have evolved beyond matrix multiplication (MatMul) to incorporate diverse non-linear operators such as GELU, SoftMax, and LayerNorm (See top box in Fig. 1). These non-MatMul operators now can account for more than 40% LLM inference latency on GPUs [1], forming new computational bottlenecks that demand both efficiency and flexibility from hardware accelerators.

As illustrated in Fig. 1, these emerging operators feature multi-stage fused pipeline kernels comprising coalesced loop bodies that are fused together by intermediate data dependencies. These kernel exhibits three key characteristics that challenge hardware design: First, the diverse loop iteration patterns per kernel stage **(Challenge 1)**, such as varying iteration counts across stages in LayerNorm, hinder regular optimization practices like loop unrolling and spatial expansion. Second, the mixed runtime data reuse strategies across stages **(Challenge 2)** require intermediate values (e.g., mean and variance) computed in earlier stages to be efficiently reused in later stages, creating complex data dependencies. Third, the evolving AI algorithm landscape demands the ability to adapt to new operator types and dataflow routines **(Challenge 3)** without requiring full hardware re-design. Thus, the complexity to support such algorithms on hardware becomes significant.

Although many heterogeneous ASIC architectures have been proposed with well-tailored sub-accelerators to tackle these challenges, their kernel-specific optimizations eliminate the ability to adapt to new operator types and dataflow routines for future algorithms. Therefore, coarse-grained reconfigurable arrays (CGRAs) become increasingly popular for their ability to combine efficiency with flexibility. With spatial programmability on the interconnect and computation at word-level granularity, spatial CGRAs deliver near-ASIC efficiency on key workloads while maintaining adaptability for emerging applications [2]. Yet, spatial CGRAs suffer from their single-command routine on each processing element (PE), forcing larger kernels to spread across different PEs until fitting in space and leading to under-utilization. Hence, it is hard for spatial CGRAs to support very large or phase-changing kernels without reconfiguration, which is the case of multi-stage non-MatMul operators.

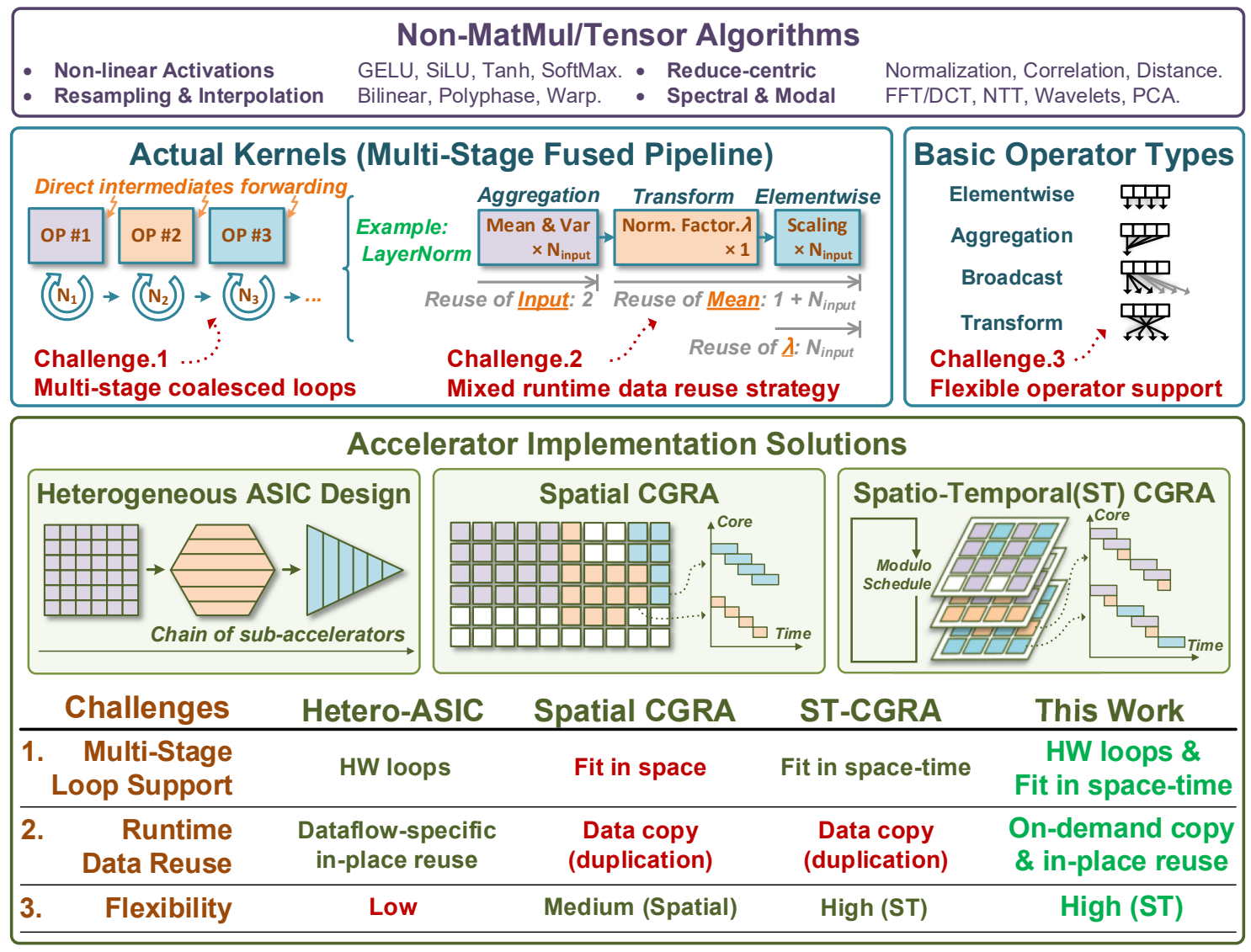

**Figure 1.** The challenge overview of supporting multi-stage fused pipeline kernels for emerging non-MatMul/Tensor algorithms in the ML-related domain: 1) multi-stage coalesced loops, 2) mixed runtime data reuse strategy, 3) flexible support of various operator types. Several limitations are observed in the existing heterogeneous ASIC design, spatial CGRA and spatio-temporal CGRA, therefore motivate the design of STELLA.

To embed such reconfigurability directly, recent spatio-temporal CGRAs (ST-CGRAs) apply the time-multiplexing strategy with modulo scheduling techniques to make each PE capable of switching its configuration on a cycle-by-cycle basis. Expanded into the time dimension, each ST-CGRA PE becomes functionally equivalent to multiple spatial CGRA PEs, enabling richer interconnects and higher flexibility. Though existing ST-CGRAs [1], [3], [4], [5] have demonstrated their ability to support complex data flow graphs with high overall utilization, the mapping of multi-stage kernels still face significant challenges. At the **application level**, when such kernels are mapped as one program, mapping inter-stage control nodes onto PEs significantly compromises the computation throughput and parallelism of the PE array. Additionally, the reuse of runtime-computed data, like the mean value and normalization factor of LayerNorm, could lead to a critical overhead of data duplication if lacking the ability to designate in-place reuse. At the **implementation level**, the additional complexity of the ST-CGRA architecture further challenges the hardware design. ST-CGRAs need to implement elastic handshake (ready-valid) logics to guarantee the synchronization of dependent PEs in the time dimension. However, these handshakes over the cyclic interconnects of the PE fabric are prone to combinational loops, thereby tightening the hardware timing headroom. Furthermore, the PE-level elastic control (e.g. auto-stall) and heavy operators (e.g. dot, non-linear) prolongs the ST-CGRA instruction routine into multiple cycles in the same time axis. Hence, a balanced system pipeline design becomes essential to absorb unnecessary delays and achieve optimal latency. As a consequence, silicon-proven ST-CGRA with low-latency support of the target multi-stage applications is scarce in the field to our knowledge.

Motivated by these limitations, this paper presents **STELLA**, an ST-elastic low-latency CGRA SoC tailored to leverage the potential of the ST-CGRA architecture and maximize the efficiency of multi-stage kernel execution. Based on the vectorized CGRA generation framework [4], STELLA comprises a 4×4 cardinal ST-CGRA fabric with 64-bit vectorized PEs, an 128 KiB L1 memory, and a RISC-V host CPU, featuring a three-fold contribution: **1)** a top-level architecture with per-PE hardware loop control, achieving **1.58-5.94x** speedups on multi-stage kernels (Sec. II); **2)** a low-latency PE design with throughput and data reuse enhancement, resolving the utilization drop in runtime data reuse scenarios with a **1.57-1.98x** speedup than the traditional CGRA solutions (Sec. III); **3)** an elastic spatio-temporal pipeline that eliminates ST-CGRA combinational loops and compacts the execution schedule (Sec. IV). Fabricated with TSMC 16nm FinFET, STELLA achieves **4.84-7.14x** better efficiency/utilization over the state-of-the-art on the multi-stage AI kernel benchmarking.

## II. STELLA SoC Architecture

As shown in Fig. 2, STELLA implements a 4x4 ST-CGRA array with a cardinal-meshed interconnect fabric, hosted by a RV32I CPU. Each CGRA PE contains a 16-deep instruction buffer (64-bit) and constant buffer (32-bit) to support modulo-scheduled execution with Initiation Interval (II) greater than one. To decouple memory access control from the precious CGRA computation resources, a dedicated Load/Store Unit (LSU) is built with self-timed address generation units (AGU). Operating on 64-bit data width, the load ports are adjacent to the top and left PEs, while store ports are at the right and bottom. First-In-First-Out (FIFO) buffers are implemented for each load/store channel to eliminate memory contentions. The host CPU configures the CGRA and LSU through Control State Registers (CSRs). To efficiently support multi-stage kernel execution, STELLA has two key features.

**A) Per-PE hardware loop control**: To address diverse loop iteration patterns across kernel stages (e.g., varying iteration counts in LayerNorm), STELLA implements a hardware loop controller (Fig. 2-①) in each PE. A 4-set context table at the CGRA top level maintains loop counts and instruction base/offset addresses for each stage. When a PE's loop controller signals completion, the next stage context is automatically dispatched, enabling seamless multi-stage execution within a single configuration. Each PE maintains its own sub-stage context, maximizing flexibility for heterogeneous loop patterns while fully utilizing the limited instruction buffer depth.

**B) Rapid configuration path**: When kernel complexity exceeds the 4-stage limit, stage partitioning requires runtime reconfiguration. STELLA employs a rapid configuration path (Fig. 2-②) that reuses the load ports during configuration. All PE instruction and constant buffers are concatenated into a 512-bit-wide parallel stream, achieving 16× bandwidth over conventional 32-bit CSR-based configuration. Instead of serving data, a double-buffered CSR manager ensures swift mode switching of the AGUs in LSU.

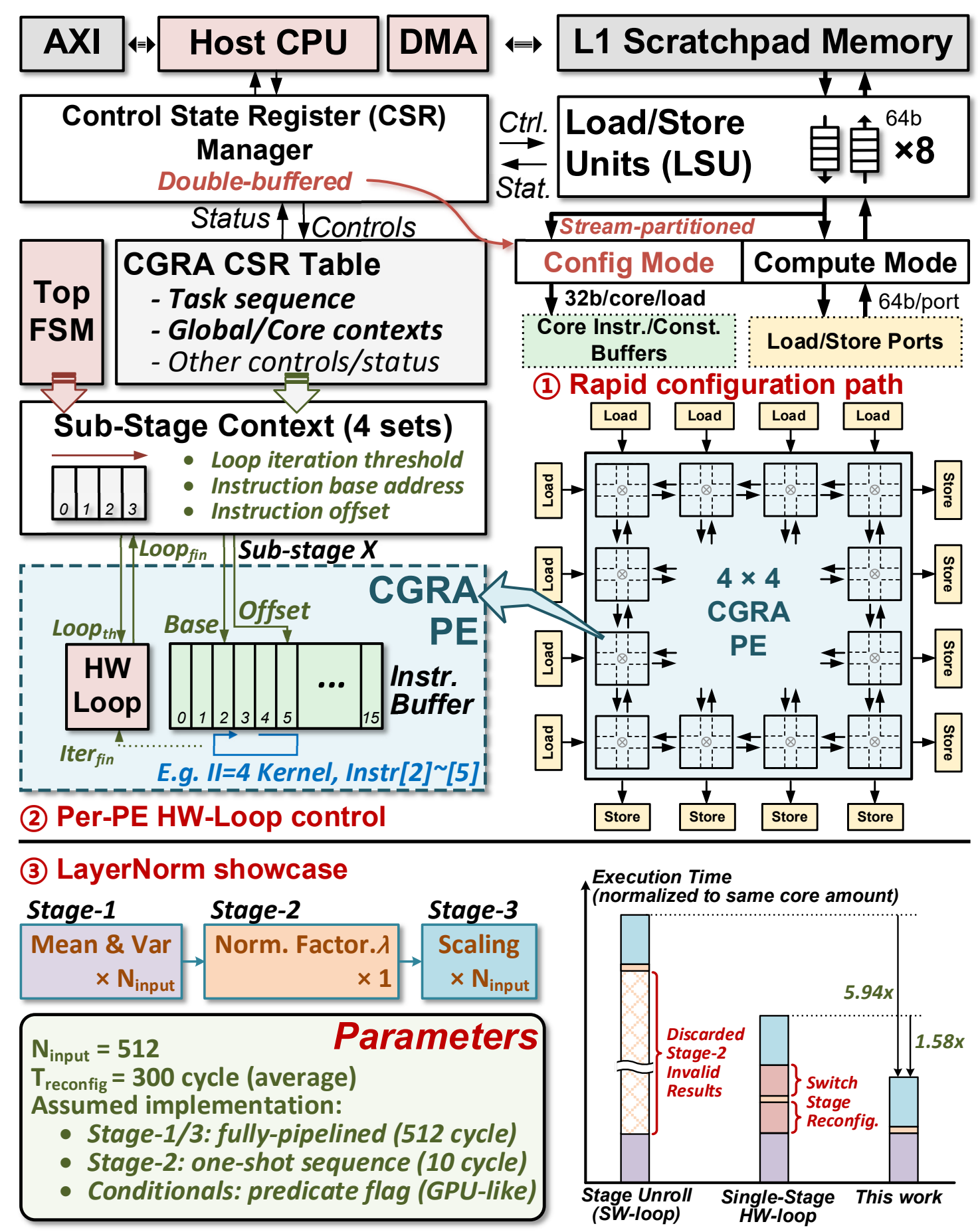

**Figure 2.** The top-level architecture of the STELLA SoC featuring a cardinal meshed 4x4 CGRA fabric with: **1)** a rapid stream-in configuration path for local instruction and constant buffers, and **2)** a per-PE hardware loop control for multi-stage kernel switching. In the showcase of 3-stage LayerNorm, the proposed architecture significantly speeds up the execution.

With these architecture add-ons, STELLA can support multi-stage coalesced loops (max. 4 stages) within a single configuration. In Fig. 2-③, a 512-point LayerNorm kernel is evaluated to compare STELLA with other solutions including the software loop (unroll 3 stages into 1 with loop controls mapped onto the CGRA PEs) and the single-stage CGRA baseline (re-configure between stages). The native multi-stage kernel support of STELLA surpasses these alternatives with 1.58-5.94× speedups as the overall computation utilization is kept maximal.

## III. Low-Latency PE Design With Runtime Data Reuse

Multi-stage fused kernels frequently require in-place reuse of intermediate results across pipeline stages. To enable efficient data reuse with minimal overhead, the STELLA PE features a vectorized SIMD datapath with three key enhancements. As detailed in Fig. 3, the main crossbar manages data routing through the PE fabric interconnects, with two runtime storage FIFO arrays: the Output FIFO buffers data for the PE fabric, while the CU Input FIFO serves as the final destination for routed data. A smaller crossbar indexes the CU FIFOs before the SIMD fixed-point CU consumes the data and executes operations. The 64-bit instruction architecture allocates control bits for the main crossbar (global router), CU input crossbar (local router), and CU operator control.

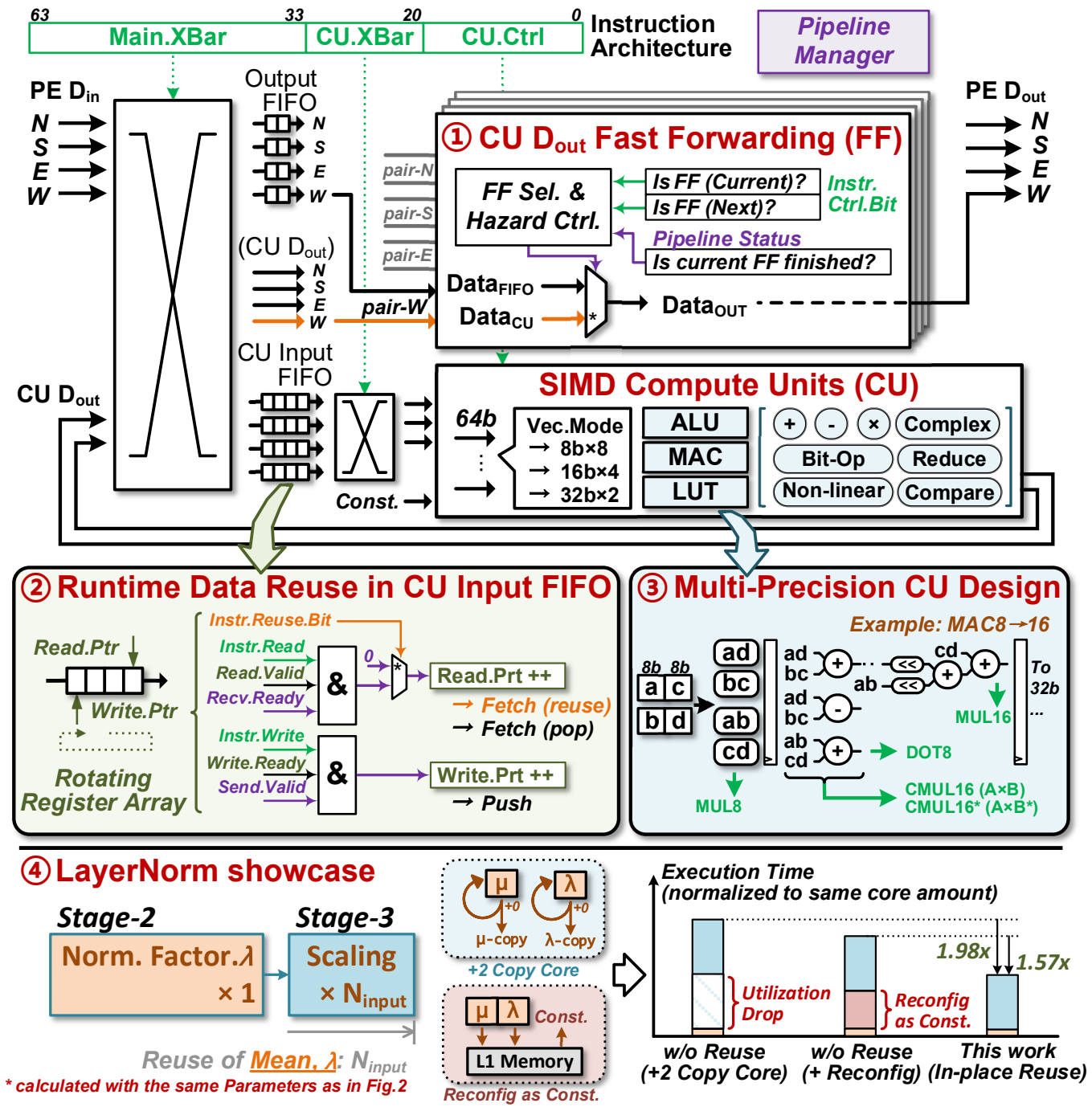

**Figure 3.** The low-latency PE design details of STELLA, including: **1)** a hazard-free fast-forwarding mechanism for compute-unit (CU) outputs to other PEs, **2)** the runtime data reuse designation with minimal control bits via CU input FIFO array, **3)** a vectorized SIMD compute unit with multi-precision support. **4)** The reuse of mean/lambda values in LayerNorm example shows the efficacy of these contributions in the in-place reuse of intermediates.

**A) Hazard-free CU output fast forwarding (FF)**: Proposed by [6], the outputs from the CU can bypass the local Output FIFO and forward directly to other PEs, saving 1 clock cycle per operation. However, the elastic timing of ST-CGRA creates hazards when sender and receiver are unsynchronized. STELLA resolves this by adding a hazard control unit (Fig. 3-①) that tracks the actual execution pipeline status, ensuring safe bypassing w/o sacrificing elasticity. The pipelining mechanism is detailed in Section. IV.

**B) Runtime Data Reuse in CU Input FIFO**: ST-CGRA favors FIFO-based data scheduling for its self-regulated temporal ordering and rate adaptation. However, FIFOs inherently oppose on-demand reuse of buffered data like depth-1 registers. Conventional workarounds such as register renaming or larger FIFO arrays incur significant control overhead and longer instruction codes. STELLA implements the FIFO using a rotating register array with read/write pointers, adding only 1 reuse bit per FIFO to enable fetch-without-pop semantics (Fig. 3-②). As shown in Fig. 3-④, this achieves data reuse with zero latency and throughput loss, outperforming alternative solutions.

**C) Multi-precision CU:** The SIMD CU (Fig. 3-③) operates on 64-bit data width with flexible vector modes (VEC8, VEC16, VEC32) selectable per instruction per PE, enabling seamless mixed-precision dataflows. Low-bit hardware is reused to construct higher-precision operators, providing native support for fixed-point complex-value operations and their conjugates at no additional cost.

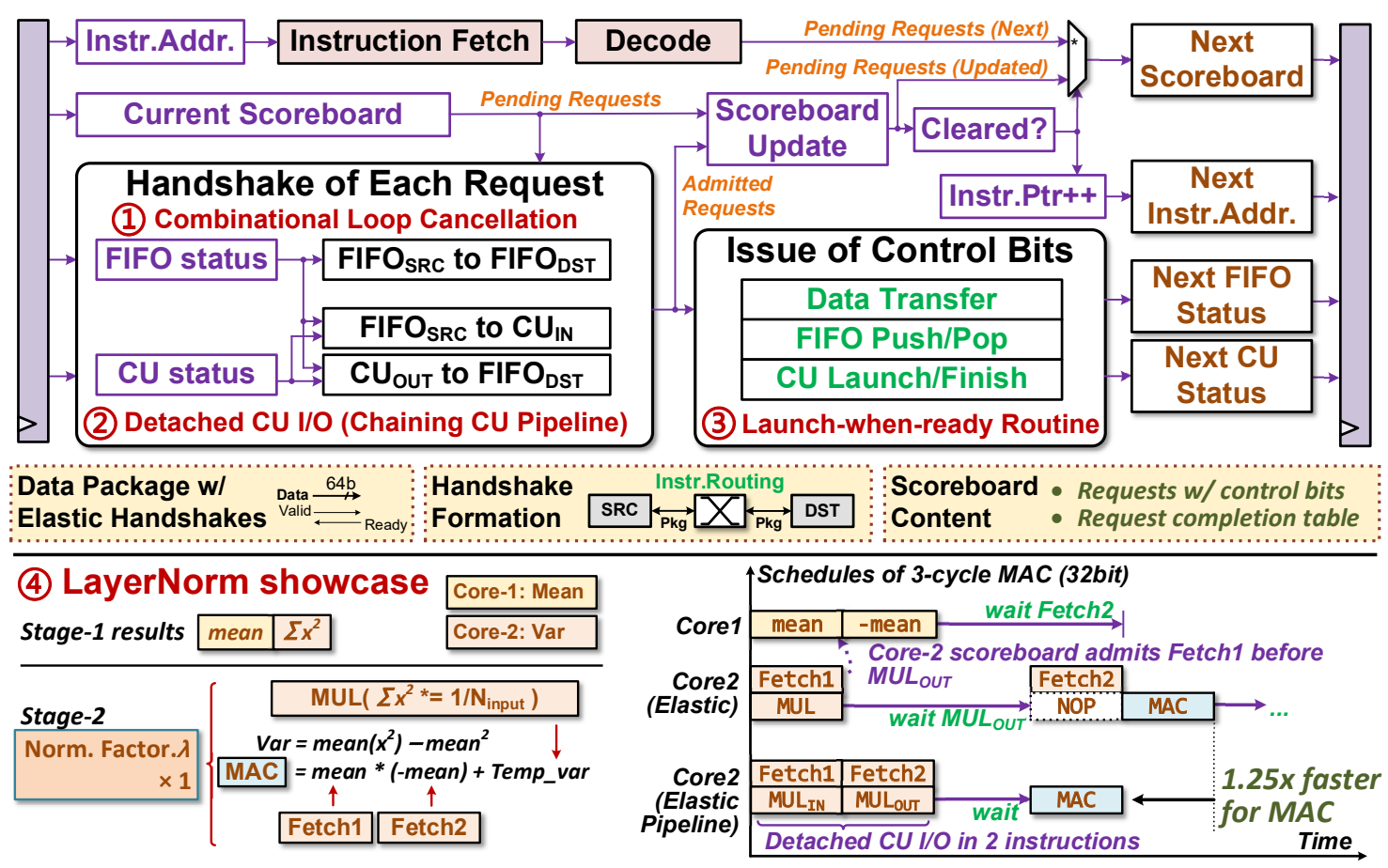

**Figure 4.** The 2-stage pipeline design of STELLA PE on top of the baseline ST-CGRA elasticity, with: **1)** the cancellation of combinational loops, **2)** the detached CU input/output handshake that chains the CU pipeline to the PE level, **3)** a scoreboard-based launch-when-ready routine to minimize the execution latency. The example of LayerNorm variance calculation achieves a more compact execution pipeline with the proposed methodology.

## IV. Elastic Low-Latency PE Pipeline

To support diverse operator types with varying latencies (1-4 cycles for ALU/MAC/LUT) and adapt to evolving dataflow routines, STELLA requires a flexible yet efficient pipeline mechanism. The challenge is compounded by the multi-operation nature of CGRA instructions, where data send/receive/bypass between PEs can incur unpredictable delays. A 2-stage elastic pipeline is proposed to resolve these issues while maintaining ST-CGRA elasticity.

**A) Spatio-Temporal elastic pipeline:** STELLA implements a 2-stage pipeline (Fig. 4-①) that decouples cyclic FIFO handshakes across the CGRA fabric. This eliminates combinational loops inherent in ST-CGRA architectures, providing larger timing margins for silicon implementation. The pipelined ready/valid handshake further decouples CU input and output requests, chaining the CU pipeline into the PE-level pipeline (Fig. 4-②). This enables deeply pipelined programs where an operator's launch and completion can be mapped to different instructions. As shown in Fig. 4-④, this optimization accelerates LayerNorm's variance computation by 20% while preserving ST elasticity.

**B) Scoreboard-based low-latency control dispatch:** To minimize false dependencies between PE subroutines constrained by the same instruction, a scoreboard (Fig. 4-③) manages pending requests and fulfills them independently when handshakes complete. This launch-when-ready mechanism enables the next instruction only after all scoreboard entries are cleared, compacting the execution schedule w/o sacrificing correctness.

With this 2-stage PE pipeline and the corresponding PE modules, STELLA is able to draw the critical path of an ST-CGRA to the same level of the spatial CGRAs. Then its well-tailored spatial-temporal flexibility and elasticity make it excel the other design on the target algorithms.

## V. Evaluation and Results

To evaluate the effectiveness of STELLA, we map several representative algorithms in the target domains, including LayerNorm,

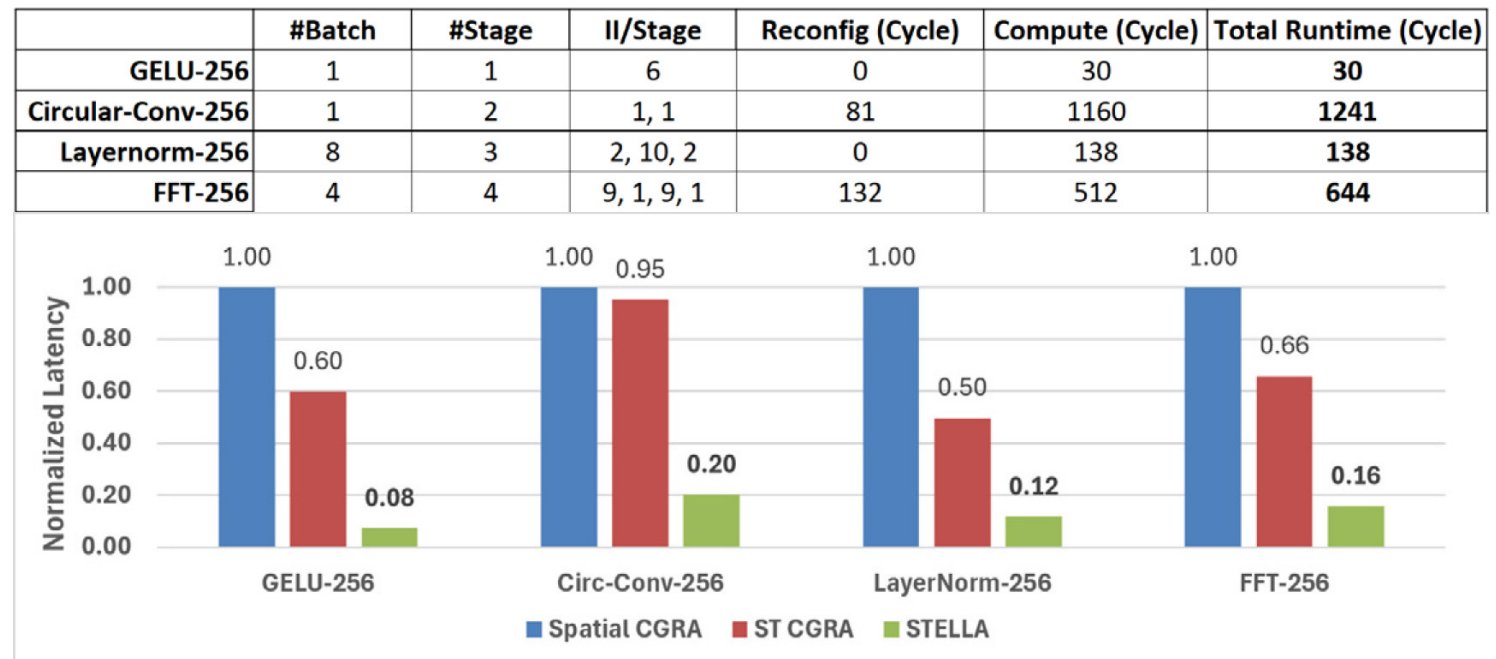

| | #Batch | #Stage | II/Stage | Reconfig (Cycle) | Compute (Cycle) | Total Runtime (Cycle) |
|---|---|---|---|---|---|---|
| GELU-256 | 1 | 1 | 6 | 0 | 30 | **30** |
| Circular-Conv-256 | 1 | 2 | 1, 1 | 81 | 1160 | **1241** |
| Layernorm-256 | 8 | 3 | 2, 10, 2 | 0 | 138 | **138** |
| FFT-256 | 4 | 4 | 9, 1, 9, 1 | 132 | 512 | **644** |

**Figure 5.** The evaluation results of mapping representative multi-stage algorithms, including GELU (polynomial-based), Circular Convolution, LayerNorm, and FFT, onto STELLA. The table shows STELLA's detailed mapping strategy and latency of each algorithm. Compared with the baseline Spatial CGRA and ST-CGRA, the speedups of STELLA are plotted in form of normalized latency overheads.

GELU, FFT, and Circular Convolution. The input size of each workload is set to 256 and the data width of computation is 16 bits. As shown in Fig. 5, the enhanced ST-CGRA architecture of STELLA helps maintain a low-latency performance among the evaluated applications with different stages of loops.

Firstly, most of these kernels result in efficient mapping with multiple instructions (II>1) to utilize the time-dimension of processing elements for time-multiplexing pipelining and cross-cycle data forwarding. Compared to this, spatial CGRAs suffer from low utilization in mapping these kernels as they have to spread the time-dimension behavior to multiple spatial processing elements, lowering the average utilization.

Secondly, a mixture of multi-cycle operators can be identified in kernels like GELU and LayerNorm, which hinders the efficiency of normal elastic ST-CGRA that inserts bubbles into the execution pipeline. Meanwhile, the rich input data reuse in GELU's polynomial approximation and LayerNorm's scaling stage costs the ST-CGRA to apply dedicated copy PEs to hold the reuse data.

As a result, STELLA achieves an average 4.84-7.14x speedup of the benchmarked application versus the baseline ST-CGRA and Spatial CGRA architecture, respectively.

## VI. Chip Measurement

STELLA is fabricated using TSMC 16nm FinFET technology, occupying 0.81 mm$^2$ for core area on a 9 mm$^2$ chip. As shown in Fig. 6, STELLA can operate up to 850 MHz@1.0V, with 325MHz@0.6V to be its most efficient point (MEP). The resource distribution chart demonstrates the overhead of local buffers and interconnects in the CGRA PE array for spatio-temporal flexibility. Yet, the proposed contributions help STELLA maintain computation efficiency. In the state-of-the-art (SotA) comparison, STELLA significantly boosts the performance of ST-CGRA thanks to the spatio-temporal PE pipeline. Powered by its vectorized and high-density CU, STELLA achieves the same level of energy efficiency against the SotA spatial CGRAs with its extra spatio-temporal flexibility.

## VII. Conclusion

This paper presents STELLA, an ST-CGRA SoC for non-MatMul/Tensor algorithms with optimized elastic and low-latency performance. Several challenges are identified during the support of kernels with multi-stage coalesced loops in the target algorithm domains. To tackle them, STELLA acquires a spatio-temporal pipelined architecture with hardware loops and on-demand in-place data reuse. Together, these innovations let STELLA achieve a 4.84-7.14x speed up across the benchmarked applications. Further, on the chip level, STELLA pushes forward the performance and efficiency boundary of ST-CGRA to the same-level of SotA spatial CGRAs, laying a promising foundation to unveil ST-CGRA potentials.

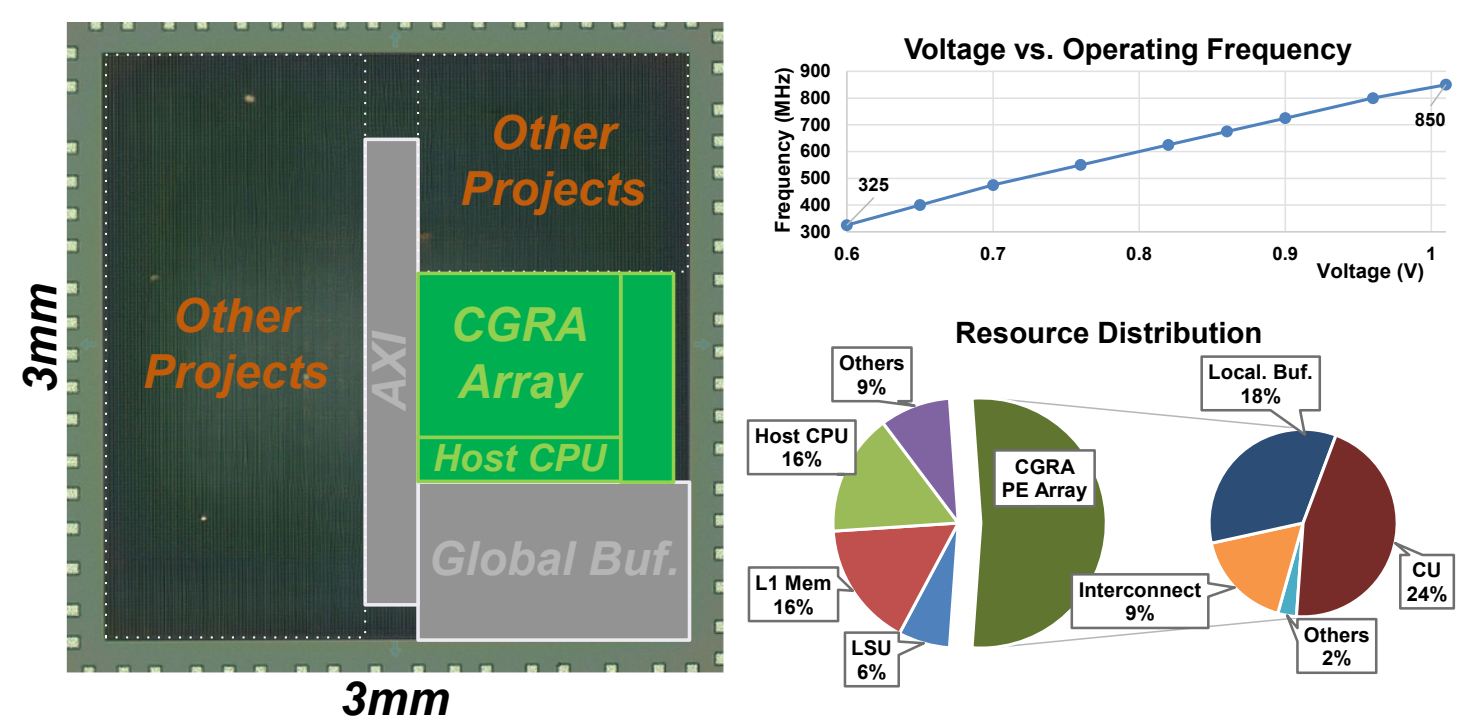

| | JSSC '23 [1] | VLSI '24 [2] | HotChips '24 [5] | STELLA |
|---|---|---|---|---|
| **Architecture** | Spatial CGRA | | **Spatio-Temporal CGRA** | |
| **Tech (nm)** | 16 | 12 | 40 | 16 |
| **Chip/Core Area (mm$^2$)** | 20.1/ 7.24 | 23/ 9.27 | 9.5/ 3.02 | 9/ 1.11 |
| **#PE** | 384 | 384 | 64 | 16 |
| **Voltage (V)** | 0.84-1.29 | 0.66-0.78 | 0.6-1.0 | 0.6-1.0 |
| **Freq. (MHz)** | 955@1.29V | 970@0.78V | 100@1.0V | 850@1.0V |
| **Peak Perf. (GOPS)** | 367@1.29V, 955MHz | 571@0.78V, 850MHz | 6.4@1.0V, 100MHz | 218@1.0V, 850MHz |
| **Energy Eff. (GOPS/W)** | 538@0.84V | 756@0.66V | 360@0.6V 154@1.0V | 555@0.6V 284@1.0V |
| **Area Eff.* (GOPS/mm$^2$)** | 50.69 | 25.99 | 33.11 | **196.4** |

**Figure 6.** The STELLA die photo, Schmoo plot, resource distribution and the comparison with state-of-the-art CGRA SoCs (INT16). The area efficiency is normalized to the 16nm technology for fair comparison.

## Acknowledgment

This project has been partly funded by the European Research Council (ERC) under grant agreement No. 101088865 and the Flanders AI Research Program. Corresponding Author: Chao Fang.